\documentclass[letterpaper]{article} 
\usepackage[preprint]{aaai2027}  
\usepackage[hyphens]{url}  
\usepackage{graphicx} 
\usepackage{natbib}  
\usepackage{caption} 
\usepackage{algorithm}
\usepackage{algorithmic}

\usepackage{newfloat}
\usepackage{listings}
\DeclareCaptionStyle{ruled}{labelfont=normalfont,labelsep=colon,strut=off} 
\floatstyle{ruled}
\newfloat{listing}{tb}{lst}{}
\floatname{listing}{Listing}

\usepackage{booktabs}

\title{ScalablePromptus: Scalable and High-Fidelity Prompt-Based Video Streaming}

\author{
    Zehao Cao\textsuperscript{\rm 1},
    Bowei Xu\textsuperscript{\rm 1},
    Xun Cao\textsuperscript{\rm 1},
    Zhan Ma\textsuperscript{\rm 1},
    Hao Chen\textsuperscript{\rm 1}\corresponding
}

\affiliations {
    \textsuperscript{\rm 1}Nanjing University\\
    caozehao22@mails.ucas.ac.cn, chenhao1210@nju.edu.cn
}

\begin{document}

\maketitle

\begin{abstract}
Prompt-based video streaming transmits compact semantic prompts instead of pixel-level content for generative reconstruction, enabling ultra-low-bitrate communication. However, the state-of-the-art Promptus framework is vulnerable to network fluctuation, where partially received prompts lead to catastrophic quality collapse. We propose ScalablePromptus, which enhances Promptus with semantic and color-aware prompt inversion, spherical linear interpolation for intermediate frames, and---most critically---a dropout training strategy that produces rank-ordered prompt representations. This allows the receiver to reconstruct meaningful video from arbitrarily truncated prompts \textbf{without any adaptation}. Under stable networks, ScalablePromptus achieves modest quality gains. Under lossy conditions, it reduces the performance degradation caused by truncation \textbf{by} $\mathbf{82\%}$--$\mathbf{95\%}$ compared to the baseline, making prompt-based streaming robust enough for real-world deployment.
\end{abstract}

\begin{links}
   \link{Code}{https://github.com/ZhChessOvO/ScalablePromptus}
\end{links}

\section{1~~~Introduction}
The rapid growth of streaming applications has led to an exponential increase in video traffic. Traditional codecs such as H.264 and H.265 compress video by removing spatial and temporal redundancies. However, since these redundancies are inherently limited, further compression forces the discarding of non-redundant content, resulting in severe artifacts such as blurring and blocking. This fundamental bound---rooted in the Shannon limit---motivates a paradigm shift: rather than compressing pixel-level signals, we can transmit compact semantic representations and reconstruct video frames at the receiver using generative models.

Promptus \cite{wu2026promptus} exemplifies this new paradigm. It inverts video frames into low-rank prompt embeddings via gradient descent and employs Stable Diffusion to regenerate frames at the receiver. Only sparse keyframe prompts are transmitted; intermediate frames are approximated through prompt interpolation. This approach achieves over $4\times$ bandwidth reduction compared to H.265 while maintaining comparable perceptual quality. Despite its promise, Promptus has two key limitations that hinder its practicality.

\begin{figure}[t!]
\centering
\includegraphics[width=1.0\columnwidth]{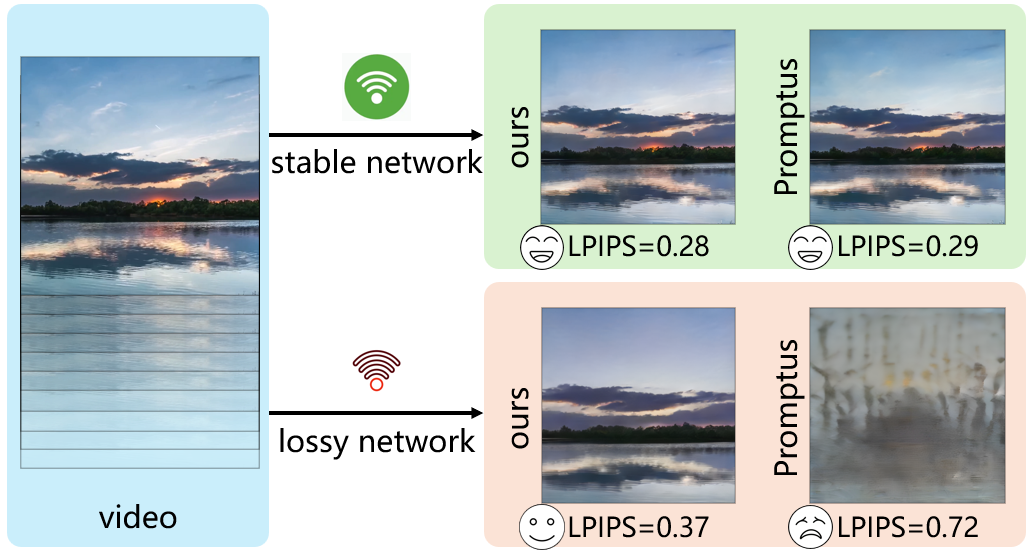}
\caption{Comparison between standard Promptus and our ScalablePromptus under different network conditions. Under ideal transmission, ScalablePromptus achieves modest but consistent quality gains through semantic/color-aware inversion and spherical interpolation. Under lossy transmission with truncation, standard Promptus suffers catastrophic quality degradation, while ScalablePromptus degrades gracefully thanks to its dropout-trained rank-ordered representations.}
\label{fig:teaser}
\end{figure}

First, the quality of the generated video is suboptimal. Promptus optimizes prompts using only pixel-level losses, which cannot fully exploit the semantic expressive capacity of the prompt space. Moreover, its linear interpolation for intermediate frames causes norm collapse in the hyperspherical embedding space, resulting in desaturated, low-contrast frames (see Sec 3.2).

Second, and most critically, Promptus lacks any mechanism to handle packet loss or unpredictable network fluctuations (see Sec 3.3). Its low-rank bitrate control requires pre-fitting multiple prompts at different ranks, which is both storage-intensive and unable to adapt to real-time bandwidth variations. Moreover, if the available bandwidth increases mid-stream, Promptus must discard the already-received low-rank prompt and re-download a higher-rank version from scratch. Conversely, if a transmitted prompt is partially lost or truncated---a common occurrence under volatile network conditions---the receiver has no way to recover meaningful video, as is shown in Figure~\ref{fig:teaser}.

To address these limitations, we propose \textbf{ScalablePromptus}, an enhanced prompt streaming framework with two contributions:

\begin{itemize}
\item \textbf{Semantic and geometric prompt enhancement.} We introduce a semantic similarity loss and a color statistics alignment loss to improve per-frame fidelity. We further replace linear interpolation with spherical linear interpolation (Slerp) during both training and inference, preserving embedding norms and eliminating the saturation degradation in intermediate frames.

\item \textbf{Dropout training for scalable streaming.} This is our core contribution. We design a dropout training strategy, where during each iteration a random subset of prompt dimensions is dropped. This forces the model to encode the most critical information in the leading dimensions, producing a rank-ordered representation. At inference time, the receiver can directly use the first $k$ dimensions of a received prompt---even if it is truncated---and still produce high-quality output. Beyond robustness to loss, this rank ordering enables a lightweight incremental quality upgrade: the base layer alone already delivers acceptable quality, and any additionally received trailing dimensions simply refine the details without requiring re-downloading or re-training.

\item We evaluate ScalablePromptus on multiple datasets. Under stable networks, the semantic and geometric enhancements yield modest but consistent quality improvements over Promptus. Under volatile conditions with truncation, the dropout training strategy reduces performance degradation \textbf{by} $\mathbf{82\%}$--$\mathbf{95\%}$, demonstrating that scalable prompt streaming is both practical and robust for real-world deployment.

\end{itemize}

\section{2~~~Related Works}

\textbf{Video streaming.} Traditional codecs \cite{1218189,6316136,9503377} compress video by eliminating spatial and temporal redundancies, but their compression ratio is bounded by the finite amount of redundancy. Deep learning-based codecs \cite{Li_2023_CVPR,Pourreza_2023_WACV,Fathima_2023_WACV,Li_2024_CVPR,LIU2024128525} replace handcrafted modules with neural networks, yet still operate within the redundancy-removal paradigm. Neural-enhanced streaming \cite{zhou2023cadmcodecawarediffusionmodeling,10.1145/3649472,10750425,11044486} discards non-redundant information and restores quality at the receiver via super-resolution or similar techniques, but suffers from domain gaps in unseen scenarios. Generative streaming \cite{Wang_2021_CVPR,li2024reparolossresilientgenerativecodec} directly synthesizes video from compact representations such as keypoints or token indices, achieving extremely low bitrates, though often restricted to specific domains like face video.

\textbf{Stable Diffusion and prompt inversion.} Stable Diffusion \cite{Rombach_2022_CVPR} generates high-quality images conditioned on text prompts through iterative denoising. Promptus \cite{wu2026promptus} is the first framework to invert video frames into pixel-aligned prompt embeddings via gradient descent, enabling SD-based video streaming. It uses single-step denoising for efficiency, low-rank decomposition for bitrate control, and interpolation-aware fitting for inter-frame compression. Our work directly builds upon and enhances Promptus.

\textbf{Semantic embedding geometry.} The prompt embeddings used in SD reside in the CLIP embedding space, which exhibits a hyperspherical structure with a well-documented modality gap between text and image embeddings~\cite{eslami2024mitigate}. Recent work has shown that naive linear interpolation in such spaces produces degenerate samples when used with diffusion models, manifesting as washed-out colors and loss of detail~\cite{landolsi2025addressing}. In parallel, prompt inversion methods for video editing \cite{Mokady_2023_CVPR} and personalization \cite{gal2022imageworthwordpersonalizing} have highlighted the importance of embedding geometry for temporal consistency. Our adoption of Slerp is grounded in these findings.

\textbf{Ordered representations via dropout.} The idea of learning ordered, progressively refined representations also appears in scalable coding and neural network pruning. Progressive neural compression \cite{Habibian_2019_ICCV} jointly optimizes layered codes that can be truncated to match bandwidth, but often requires separate training for each layer. Nested Dropout was originally proposed for representation learning \cite{rippel2014learning} and later applied to variational autoencoders for ordered latent spaces \cite{10049079}. More recently, \citet{11045090} introduced priority-based dropout for image transmission over noisy channels, but their method is not generative. In the prompt-based streaming context, our key novelty is to combine rank-ordered prompt factors with spherical interpolation and semantic losses, enabling a single model to support arbitrary rank truncation without retraining---an advantage not present in prior scalable coding or dropout works.

\begin{figure*}[t]
\centering
\includegraphics[width=1.0\textwidth]{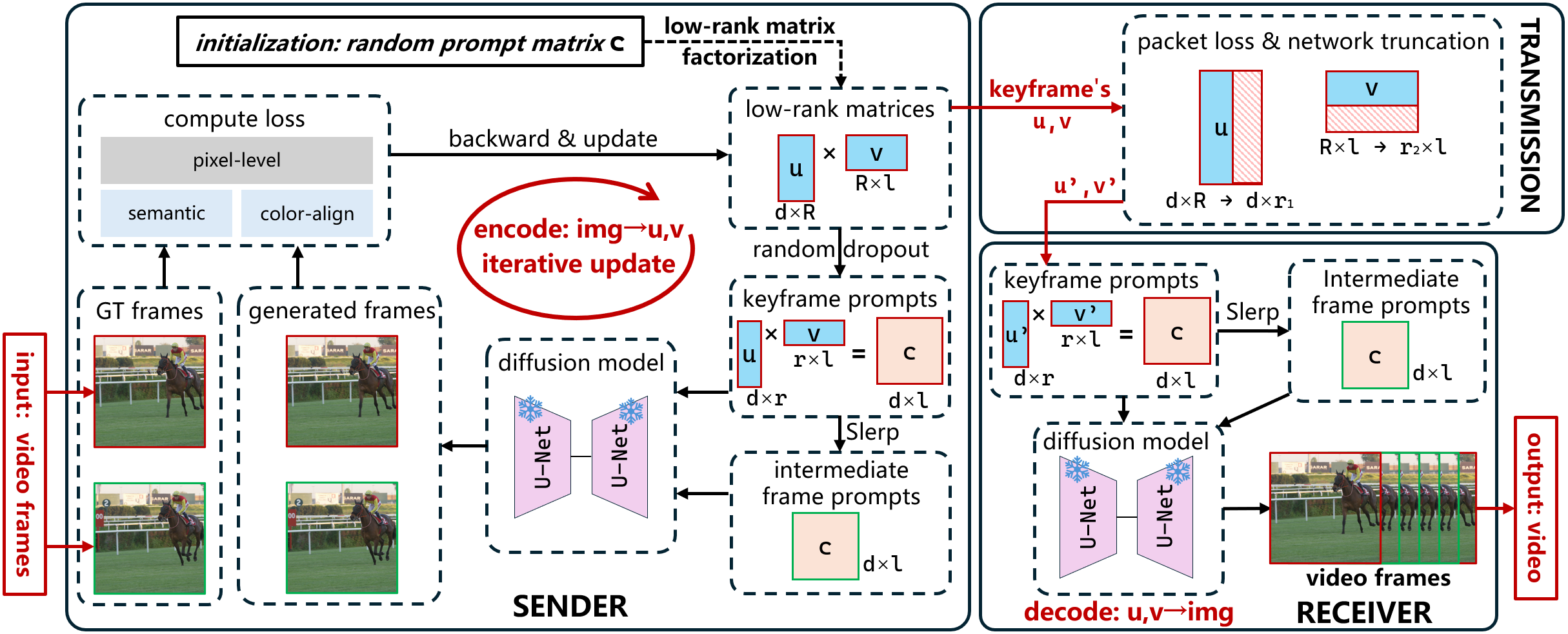}
\caption{The ScalablePromptus pipeline. At the sender, video frames are inverted into rank-ordered low-rank prompt factors via gradient descent with semantic and color constraints, and trained with dropout. During transmission, only keyframe prompt matrices are sent; packet loss and network truncation may make the received prompts to a lower effective rank. At the receiver, spherical interpolation recovers intermediate frames, and Stable Diffusion reconstructs the video from the prompts.}
\label{fig:pipeline}
\end{figure*}

\section{3~~~Method}
In this section, we present ScalablePromptus, our enhanced prompt streaming framework. We first provide an overview of the pipeline and then detail the two key components: semantic and geometric prompt enhancement, and dropout training for scalable streaming.

\subsection{3.1 Overview of ScalablePromptus}
Figure~\ref{fig:pipeline} provides an overview of the ScalablePromptus framework. The pipeline consists of three stages: sender-side prompt inversion, transmission, and receiver-side video generation.

\textbf{Sender side.} The input video is first divided into segments bounded by scene change detection. Within each segment, keyframes are selected at a fixed interval. Each keyframe is inverted into a low-rank prompt representation $c = U \cdot V$ through gradient descent, where $U \in \mathbf{R}^{d \times r}$ and $V \in \mathbf{R}^{r \times l}$ are the low-rank factors with rank $r$. Unlike the original Promptus, our inversion optimizes a composite loss that jointly considers pixel-level fidelity, semantic similarity, and color statistics alignment. Moreover, during training, intermediate frames are obtained via spherical linear interpolation (Slerp) rather than linear interpolation; gradients flow through the interpolation back to the keyframe prompts, ensuring that the optimized keyframes produce optimal intermediate frames under Slerp.

Additionally, we train the prompt factors with a dropout method. This forces the model to encode the most essential visual information in the leading dimensions. As a result, the rank indices are ordered by importance: earlier dimensions carry coarse, critical content, while later dimensions encode fine-grained details. We refer to this property as \textbf{rank-ordered representation}.

\textbf{Transmission.} Only the prompt factors of keyframes are transmitted. To further reduce bitrate, both $U$ and $V$ are quantized to 8-bit integers before transmission. The prompts for non-keyframes are not transmitted; they are recovered at the receiver via interpolation. Under ideal network conditions, the full-rank prompt factors are delivered intact. However, in practical streaming scenarios, packet loss or bandwidth drops may cause loss events that render not only the lost data but also all subsequent dependent data unusable. With standard Promptus, such truncation would lead to catastrophic quality degradation (see Sec 4.2). Thanks to the rank-ordered property induced by dropout, ScalablePromptus can tolerate this gracefully.

\textbf{Receiver side.} Upon receiving the (possibly truncated) keyframe prompts, the receiver first reconstructs the prompt embeddings: $c = U_{:, :k} \cdot V_{:k, :} \;/\; \sqrt{k}$, where $k$ is the number of successfully received dimensions. For intermediate frames between keyframes, we apply spherical linear interpolation (Slerp) in the prompt space. Finally, the interpolated prompts are fed into Stable Diffusion along with the noised previous frame to generate each video frame. The generated frame is then used as the basis for the next frame's noise input, maintaining temporal continuity throughout the sequence.

\subsection{3.2 Semantic and Geometric Prompt Enhancement}
\label{sec:semantic-geometric}
The original Promptus framework optimizes prompt embeddings using pixel-level losses: a weighted combination of LPIPS perceptual loss and MSE reconstruction loss, plus a prompt regularization term. While effective for fitting individual frames, these losses operate purely at the pixel level and cannot fully exploit the semantic expressive capacity of the prompt space. Moreover, Promptus approximates intermediate frames via linear interpolation, which causes norm collapse in the hyperspherical embedding space. We address both issues through two loss constraints and spherical interpolation, as detailed below.

\subsubsection{Semantic Similarity Loss}

Prompts in Stable Diffusion condition the generation through the CLIP text encoder; thus, the prompt embedding space is inherently semantic. However, the original Promptus loss measures only pixel-level discrepancy, which does not guarantee semantic alignment. To enforce this, we introduce a cosine similarity loss between their semantic embeddings.

We load a pretrained CLIP ViT-B/32 image encoder in evaluation mode with frozen weights to serve as the semantic feature extractor. Let $\phi(\cdot)$ denote the embedding extracted by this encoder. The semantic similarity loss between a generated frame $\hat{x}$ and its ground truth $x$ is:

\begin{equation}
\mathcal{L}_{\mathrm{sem}} = 1 - \frac{\phi(\hat{x}) \cdot \phi(x)}{\|\phi(\hat{x})\| \; \|\phi(x)\|}.
\end{equation}

The image preprocessing pipeline is implemented entirely in PyTorch and is fully differentiable, allowing gradients to flow through the encoder back to the prompt embeddings.

\subsubsection{Color Statistics Alignment Loss}

Frames generated by Promptus sometimes exhibit global color shifts relative to the ground truth (e.g., appearing grayer, see Figure 6 "only semantic"). Since color perception is largely governed by channel-wise statistics, we propose a color statistics alignment loss that matches per-channel RGB means and standard deviations.

Let $\mu(\hat{x})_c$ and $\sigma(\hat{x})_c$ denote the mean and standard deviation of the generated frame $\hat{x}$ in channel $c \in \{R, G, B\}$, and similarly $\mu(x)_c$ and $\sigma(x)_c$ for the ground-truth frame $x$. The color loss is:

\begin{equation}
\mathcal{L}_{\mathrm{color}} = \frac{1}{3} \sum_{c \in \{R,G,B\}} \Big[ \big(\mu(\hat{x})_c - \mu(x)_c\big)^2 + \big(\sigma(\hat{x})_c - \sigma(x)_c\big)^2 \Big].
\end{equation}

This loss corrects systematic color biases without affecting the fine-grained structure captured by other losses.

\subsubsection{Overall Composite Loss}

Combining all components, the complete loss function is:

\begin{equation}
\begin{array}{l}
\mathcal{L} = \underbrace{\alpha \cdot \mathcal{L}_{\mathrm{LPIPS}} + \beta \cdot \mathcal{L}_{\mathrm{MSE}} + \lambda_{\mathrm{reg}} \cdot \|c - \mu\|^2}_{\mbox{base Promptus loss}} \\
\qquad + \; \underbrace{w_{\mathrm{sem}} \cdot \mathcal{L}_{\mathrm{sem}} + w_{\mathrm{color}} \cdot \mathcal{L}_{\mathrm{color}}}_{\mbox{proposed constraints}},
\end{array}
\end{equation}

where $\alpha$ and $\beta$ balance perceptual and reconstruction fidelity, $\lambda_{\mathrm{reg}}$ prevents prompt drift, and $w_{\mathrm{sem}}$, $w_{\mathrm{color}}$ control the two proposed constraints.

\subsubsection{Spherical Interpolation for Intermediate Frames}

\begin{figure}[t]
\centering
\includegraphics[width=1.0\columnwidth]{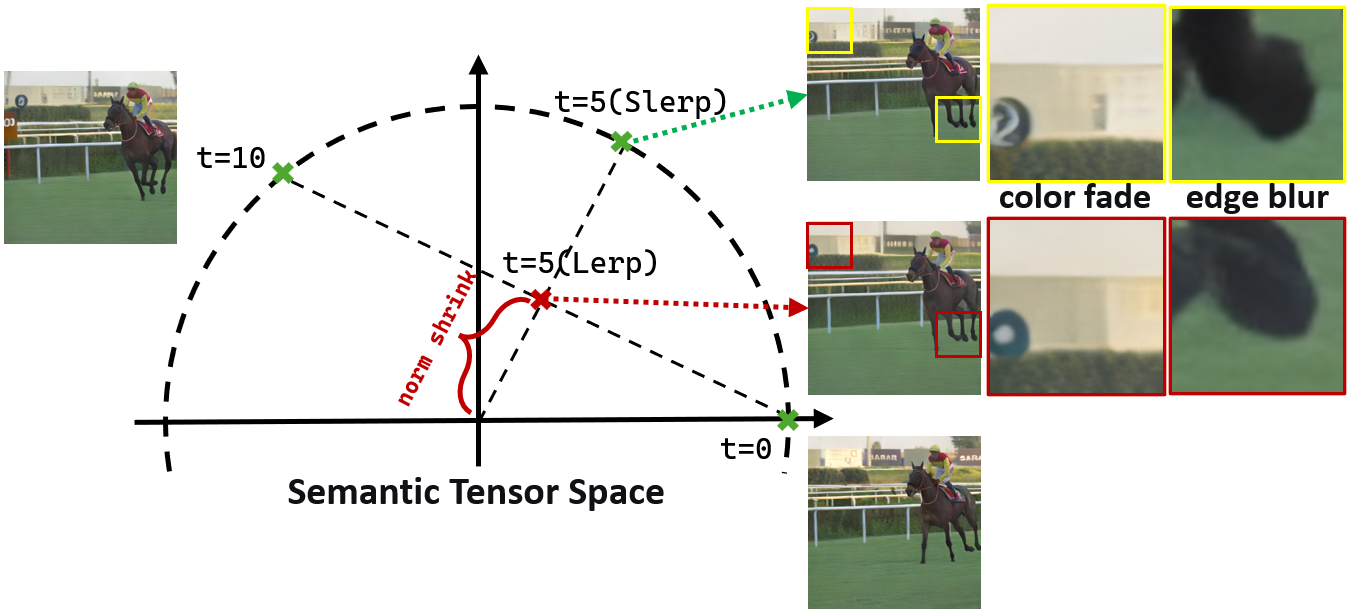}
\caption{Illustration of linear interpolation (Lerp) and spherical linear interpolation (Slerp).}
\label{fig:slerp}
\end{figure}

Promptus transmits only sparse keyframe prompts and approximates intermediate frames via linear interpolation (Lerp) in the prompt space. However, the prompt embeddings reside on a hyperspherical manifold~\cite{eslami2024mitigate}, and linear interpolation cuts a chord through the interior of the sphere, causing the norm of embeddings to shrink (norm collapse). This weakens the conditioning signal passed to Stable Diffusion, resulting in desaturated colors and loss of detail~\cite{landolsi2025addressing}, as illustrated in Figure~\ref{fig:slerp}. In the Lerp result, the sky appears noticeably darker and dull yellow, the logo on the building is blurred into an indistinct patch, and the horse's feet are fused together with unclear boundaries. Slerp restores natural sky tones, sharpens structural details, and produces cleanly separated contours.

We replace Lerp with spherical linear interpolation (Slerp), which travels along the geodesic on the hypersphere. Given two prompt embeddings $c_1$ and $c_2$ from adjacent keyframes, and an interpolation parameter $\omega \in [0, 1]$, Slerp is defined as:

\begin{equation}
\begin{array}{l}
\mathrm{Slerp}(c_1, c_2; \omega) = \frac{\sin\big((1 - \omega) \cdot \theta\big)}{\sin\theta} \cdot c_1 + \frac{\sin(\omega \cdot \theta)}{\sin\theta} \cdot c_2, \\[4pt]
\quad \theta = \arccos\!\left(\frac{c_1 \cdot c_2}{\|c_1\| \; \|c_2\|}\right).
\end{array}
\end{equation}

When $\|c_1\| = \|c_2\|$, Slerp preserves the embedding norm for all $\omega$, i.e., $\|\mathrm{Slerp}(c_1, c_2; \omega)\| = \|c_1\|$. This ensures every intermediate frame receives the same conditioning strength as the keyframes.

Slerp is integrated into both training and inference. During training, intermediate-frame prompts are computed via Slerp from learnable keyframe prompts, and the composite loss (Equation~4) is evaluated on all frames. Gradients flow through Slerp back to the keyframe prompts, ensuring optimality \textit{under spherical interpolation}. At inference, the receiver applies the same Slerp operation for consistency.

\subsection{3.3 Dropout Training for Scalable and Resilient Streaming}
\label{sec:nested-dropout}

The two preceding components improve prompt inversion quality under ideal transmission. However, as argued in the introduction, the most critical limitation of Promptus is its inability to handle packet loss and bandwidth fluctuation. In this section, we present the core of ScalablePromptus: a dropout-based training strategy that produces \textit{rank-ordered} prompt representations, enabling graceful quality degradation under arbitrary truncation.

\subsubsection{Motivation: The Vulnerability of Fixed-Rank Prompts}

In standard Promptus, each keyframe prompt is trained at a fixed rank $r$. At transmission time, the sender selects among pre-fitted prompts of different ranks according to the estimated bandwidth. This approach has three fundamental weaknesses. First, it requires storing multiple copies of each prompt at different ranks, incurring significant storage overhead. Second, it cannot adapt to time-varying bandwidth: if the available bandwidth increases mid-stream, the only way to upgrade quality is to discard the already-received low-rank prompt and re-download a higher-rank version entirely, wasting both bandwidth and the previously transmitted data. Third, and most critically, it cannot react to real-time packet loss: if a rank-$r$ prompt is partially lost during transmission, the receiver simply cannot use it, because the prompt was never trained to be meaningful at any rank other than $r$. Our experiments confirm the severity of this vulnerability. This catastrophic situation makes standard Promptus impractical for any deployment scenario involving lossy networks.

In contrast, our rank-ordered representation enables a lightweight quality upgrade mechanism. The receiver starts by receiving the first $r_{\min}$ dimensions, which already carry the most critical visual information and produce acceptable quality. If additional bandwidth becomes available, the sender simply transmits the next few trailing dimensions as a supplement; the receiver appends them to the existing prompt without discarding any previously received data. This incremental refinement continues seamlessly up to the full rank $r$, making ScalablePromptus naturally suited for dynamic networks with fluctuating bandwidth.

\subsubsection{Dropout Training Strategy}

To address this, we design a training strategy inspired by Nested Dropout~\cite{rippel2014learning}. The key insight is to train the model to expect truncation, so that it learns to encode information in a priority order. Algorithm~\ref{alg:dropout} formalizes the training procedure.


\begin{algorithm}[tb]
\caption{Dropout Training for Rank-Ordered Prompts}
\label{alg:dropout}
\textbf{Input}: Keyframe set $\mathcal{K}$, max rank $r$, min rank $r_{\min}$, warmup iterations $T_w$, total iterations $T$\\
\textbf{Output}: Prompt factors $U \in \mathbf{R}^{d \times r}$, $V \in \mathbf{R}^{r \times l}$\\
\begin{algorithmic}[1]
\STATE Initialize $U, V$ randomly.
\FOR{$t = 1$ \TO $T$}
    \IF{$t \leq T_w$}
        \STATE $k \gets r$ \COMMENT{Warmup: full rank}
    \ELSE
        \STATE $k \gets \mathrm{RandomInteger}(r_{\min}, r)$
    \ENDIF
    \STATE Truncate: $U_k \gets U_{:, 1:k}$, $V_k \gets V_{1:k, :}$.
    \STATE Reconstruct prompt: $c_k \gets U_k \cdot V_k \;/\; \sqrt{k}$.
    \STATE Compute Slerp-interpolated prompts for intermediate frames from $\{c_k\}$.
    \STATE Generate frames via Stable Diffusion; compute $\mathcal{L}$ using Equation~4.
    \STATE Back-propagate gradients only through $U_k$, $V_k$.
    \STATE Apply auxiliary regularization on trailing dimensions:
    \STATE \quad $\mathcal{L}_{\mathrm{aux}} \gets \eta \cdot \big( \|U_{:, k+1:r}\|_F + \|V_{k+1:r, :}\|_F \big)$.
    \STATE $\mathcal{L}_{\mathrm{total}} \gets \mathcal{L} + \mathcal{L}_{\mathrm{aux}}$.
    \STATE Update $U$, $V$ via gradient descent.
\ENDFOR
\STATE \textbf{return} $U$, $V$.
\end{algorithmic}
\end{algorithm}


This mechanism induces a natural priority order: dimensions with lower indices are forced to recover a reasonable result alone, while higher-index dimensions are trained to contain supplementary information like texture details. Consequently, the model learns to encode the most critical, coarse visual information in the leading dimensions, and progressively finer details in the trailing dimensions. We refer to this property as \textbf{rank-ordered representation}.

\subsubsection{Training Techniques for Stability}

Training with random truncation introduces several challenges that require careful handling.

\textbf{Warmup.} At the start of training, $U$ and $V$ are initialized randomly. If dropout is enabled immediately, only the first few dimensions receive gradients, while the rest remain untrained. To ensure all dimensions are initialized with meaningful values, we use full-rank training for the first $T_w$ iterations (lines~3--5). We set $T_w = 500$ for the first frame and $T_w = 100$ for subsequent frames.

\textbf{Auxiliary regularization.} Even after warmup, the trailing dimensions are naturally trained less frequently than the leading ones, and can gradually drift or become dormant. To prevent this, we add a small auxiliary loss that penalizes large magnitudes in the non-selected dimensions (line~11), with a small coefficient $\eta = 10^{-3}$. This keeps the trailing dimensions near zero when not in use, preventing them from introducing noise when they are eventually selected.

\textbf{Smoothed worst-step selection.} Promptus dynamically allocates extra training iterations to the frame with the highest current loss. However, with random truncation, a frame may show a high loss not because it is inherently difficult, but because it was unluckily assigned a small $k$. To avoid this bias, we replace the raw loss with an exponential moving average (EMA) with decay factor $\gamma = 0.7$ when selecting the worst step. This smooths out the stochasticity introduced by sampling $k$.

\subsubsection{Inference with Arbitrary Truncation}

At inference time, the receiver may receive only a usable prefix of the transmitted $U$ and $V$ factors, because a loss event renders the lost data and all subsequent dependent data unusable. Let $k_{\mathrm{rx}}$ be the number of successfully received columns of $U$ and rows of $V$. The receiver simply reconstructs the prompt as $c = U_{:, :k_{\mathrm{rx}}} \!\cdot V_{:k_{\mathrm{rx}}, :} \;/\; \sqrt{k_{\mathrm{rx}}}$ and proceeds with the standard Slerp-based frame generation.

No communication is needed between sender and receiver regarding the truncation level, and no retraining or fine-tuning is required. The rank-ordered property ensures that the first $k_{\mathrm{rx}}$ dimensions contain the most important visual information, and the quality degrades only gradually as $k_{\mathrm{rx}}$ decreases.

\section{4~~~Experiments}

\begin{figure*}[t!]
\centering
\includegraphics[width=0.92\textwidth]{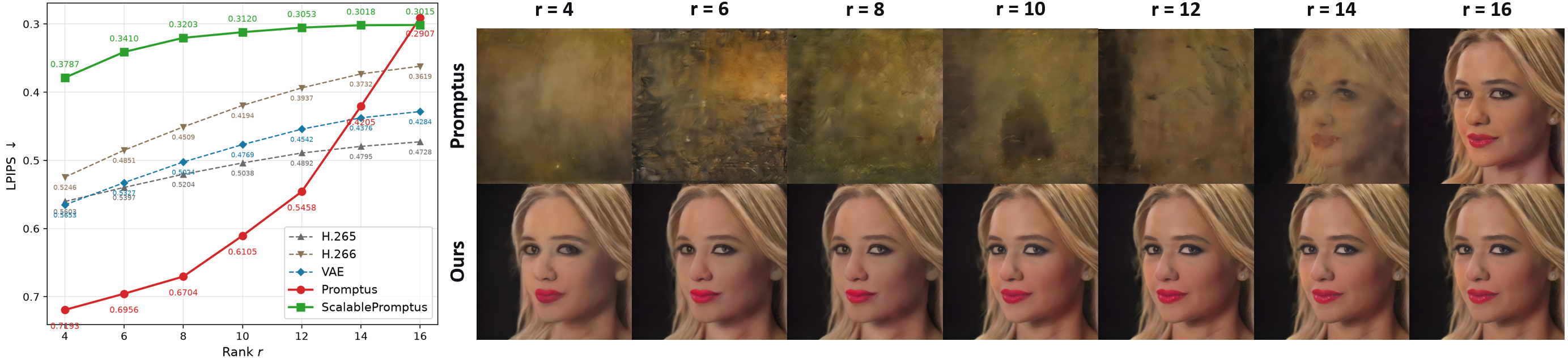}
\caption{Scalability experiments. Left: LPIPS vs.~rank. For comparison, we also include results of non-prompt methods under same transmission bitrates (encoded at 140 kbps, i.e. $r = 16$, but transmitted at low bitrates by dropping P frames). Right: visual examples at selected ranks. Our method remains resonable results under severe truncations.}
\label{fig:scalable}
\end{figure*}

\subsection{4.1 Experimental Setup}

\subsubsection{Datasets}
We evaluate on two video datasets with diverse content: QST \cite{10.1007/978-3-030-58558-7_18} and UVG \cite{10.1145/3339825.3394937}. The domains of these videos span natural landscapes and human activities, outdoor and indoor scenes. All videos are center-cropped and resized to $512 \times 512$ at 30 fps. \textbf{We adopt this resolution to maintain a fair comparison with non-prompt-based codecs}, whose bitrate scales with resolution. In contrast, prompt-based methods transmit prompts of a fixed size determined solely by the rank, regardless of the output resolution. \textbf{The receiver can generate video at any resolution by simply specifying a larger output, meaning that the compression ratio of prompt-based streaming grows with resolution at no additional bandwidth cost} \cite{wu2026promptus}.

\subsubsection{Baselines}
We compare against \textbf{H.265} \cite{6316136} and \textbf{H.266} \cite{9503377} (traditional video codecs), \textbf{VAE-based} compression \cite{Rombach_2022_CVPR} (an autoencoder, same handle as Promptus experiments), and \textbf{Promptus} \cite{wu2026promptus} (the state of the art and the original prompt streaming framework).

\subsubsection{Implementation Details}
We use Stable Diffusion v1.5 with single-step denoising, following the same generation backbone as Promptus. The prompt dimensions are $d = 1024$, $l = 77$. Low-rank factors $U$ and $V$ are quantized to 8-bit integers before transmission. For our semantic constraints, we employ a frozen CLIP ViT-B/32 as the feature extractor $\phi(\cdot)$. Dropout training uses $r_{\max} = 16$ and $r_{\min} = 4$. Warmup iterations are $T_w = 500$ for the first frame and $T_w = 100$ for subsequent frames. The total training iteration count is $10000$ for the first frame and $1500$ for subsequent frames. Default loss weights are set as $\alpha = 0.2$, $\beta = 0.8$, $\lambda_{\mathrm{reg}} = 0.1$, $w_{\mathrm{sem}} = 0.5$, and $w_{\mathrm{color}} = 0.3$. The keyframe interval is set to 10 frames (3 keyframes per second) unless otherwise specified.

\subsubsection{Complexity}
Training speed is approximately 150 ms per iteration on a single NVIDIA RTX 4090 GPU, compared to 130 ms for the original Promptus. On the receiver side, with TAESD and TensorRT acceleration, our framework achieves a decoding speed of 160 FPS, which is more than sufficient for on-demand streaming and real-time playback scenarios.

\subsection{4.2 Scalability: Performance Under Rank Truncation}

A defining advantage of ScalablePromptus is its ability to serve multiple effective ranks from a single trained model, and to degrade gracefully when prompts are truncated. We train a single set of prompt factors at $r = 16$ with dropout, and evaluate by truncating to $k \in \{4, 6, 8, 10, 12, 14, 16\}$ at inference time.

Figure~\ref{fig:scalable} presents both the quantitative LPIPS curves and visual examples at selected ranks. Our dropout-trained model maintains competitive LPIPS across all ranks: quality degrades gradually as $k$ decreases, and even at $k = 4$ the output remains structurally meaningful. In contrast, naive truncation of a standard Promptus prompt causes a catastrophic quality collapse once the effective rank falls below the trained rank---the generated frames quickly become unrecognizable. Quantitatively, our method suppresses the LPIPS degradation increment \textbf{by} $\mathbf{82\%}$--$\mathbf{95\%}$ under rank reduction, corresponding to an absolute degradation reduction of approximately $\mathbf{0.35}$--$\mathbf{0.37}$. This demonstrates that when the effective bandwidth deteriorates sharply, the baseline suffers a complete quality breakdown, while ours maintains near-stable performance, exhibiting strong robustness.

Beyond quality, our rank-ordered representation enables a practical advantage unique to ScalablePromptus: \textbf{lightweight incremental quality upgrade}. Because critical information is concentrated in the leading dimensions, the base layer alone (e.g., $k = 4$) already produces acceptable visual quality. If additional bandwidth is available mid-stream, the sender simply transmits the previously unsent trailing dimensions as a supplement; the receiver appends them to the existing prompt without any re-downloading or re-training. In contrast, standard Promptus must discard the already-received low-rank prompt and re-download a higher-rank version from scratch, multiplying the transmission, storage, and training cost by the number of supported bitrate levels. ScalablePromptus requires only a single dropout-trained model per keyframe and supports any rank on demand, making it ideally suited for dynamic network environments with unpredictable bandwidth fluctuation.


\subsection{4.3 Packet Loss Resilience}

We further evaluate robustness under realistic lossy conditions using the Gilbert-Elliott \cite{6769369} burst loss model, which captures the temporal correlation of packet losses typical of wireless networks. We configure the GE model with transition probabilities that yield average loss rates of $0\%$, $5\%$, $10\%$, and $20\%$. We set transition probabilities $p_{\mathrm{gb}} = 0.1$ and $p_{\mathrm{bg}} = 0.3$. The lost dimensions are simply set to zero before reconstruction, and both operate under the same total bandwidth of 140 kbps ($r = 16$). Table~\ref{tab:loss} shows the results. \textbf{Additional channel tests and analyses are provided in the supplementary material.}

\begin{table}[t]
\centering
\caption{LPIPS under Gilbert-Elliott packet loss.}
\label{tab:loss}
\begin{tabular}{lcccc}
\toprule
Method & $0\%$ loss & $5\%$ loss & $10\%$ loss & $20\%$ loss \\
\midrule
Promptus  & \textbf{0.2907} & 0.3345 & 0.3607 & 0.4491 \\
Ours & 0.3015 & \textbf{0.3214} & \textbf{0.3276} & \textbf{0.3417} \\
\bottomrule
\end{tabular}
\end{table}

\subsection{4.4 Comparison Under Stable Networks}

We evaluate under ideal, lossless network conditions where the sender can pre-select the optimal rank for the available bandwidth. For each method, we test at multiple bitrates by varying the codec quantization parameter (H.265, H.266, VAE) or the prompt rank (Promptus and our variants).

Table~\ref{tab:stable} reports LPIPS, masked-SSIM, and masked-PSNR (mask active regions, focusing evaluation on the challenging dynamic content where quality differences emerge) at representative bitrates. Two key observations emerge. First, Ours w/o dropout consistently outperforms the original Promptus across all bitrates, validating the benefits of semantic/color-aware inversion and spherical interpolation. Second, Ours (full), which incorporates dropout training, performs comparably to Promptus at the low bitrate and slightly trails it at the medium bitrate. This is expected: dropout training forces the model to sacrifice some representational capacity at the trained rank in exchange for robustness at lower ranks. As demonstrated in Sec~4.2 and~4.3, this minor degradation under ideal conditions is more than compensated by dramatic improvements under truncated and lossy scenarios. \textbf{See detailed analyses in supplementary material.}

\begin{table}[t]
\centering
\caption{Quantitative comparison under stable networks.}
\label{tab:stable}
\begin{tabular}{lccc}
\toprule
Method  & LPIPS$\downarrow$ & mSSIM$\uparrow$ & mPSNR$\uparrow$ \\
\midrule
\multicolumn{4}{c}{\textit{Low bitrate (70 kbps)}} \\
H.265   & 0.5017 & 0.4837 & 22.51 \\
H.266   & 0.4361 & 0.5167 & 23.05 \\
VAE     & 0.4862 & 0.4929 & 22.64 \\
Promptus    & 0.3476 & 0.5813 & 23.43 \\
Ours w/o dropout & \textbf{0.3075} & \textbf{0.6820} & \textbf{23.84} \\
Ours (full) & 0.3355 & 0.6447 & 22.80 \\
\midrule
\multicolumn{4}{c}{\textit{Medium bitrate (140 kbps)}} \\
H.265   & 0.4728 & 0.5640 & 23.70 \\
H.266   & 0.3619 & 0.6099 & 24.44 \\
VAE     & 0.4284 & 0.5837 & 24.00 \\
Promptus    & 0.2950 & 0.6935 & 24.32 \\
Ours w/o dropout & \textbf{0.2875} & \textbf{0.6959} & \textbf{24.63} \\
Ours (full) & 0.3137 & 0.6578 & 23.55 \\
\bottomrule
\end{tabular}
\end{table}

\begin{figure}[t]
\centering
\includegraphics[width=1.0\columnwidth]{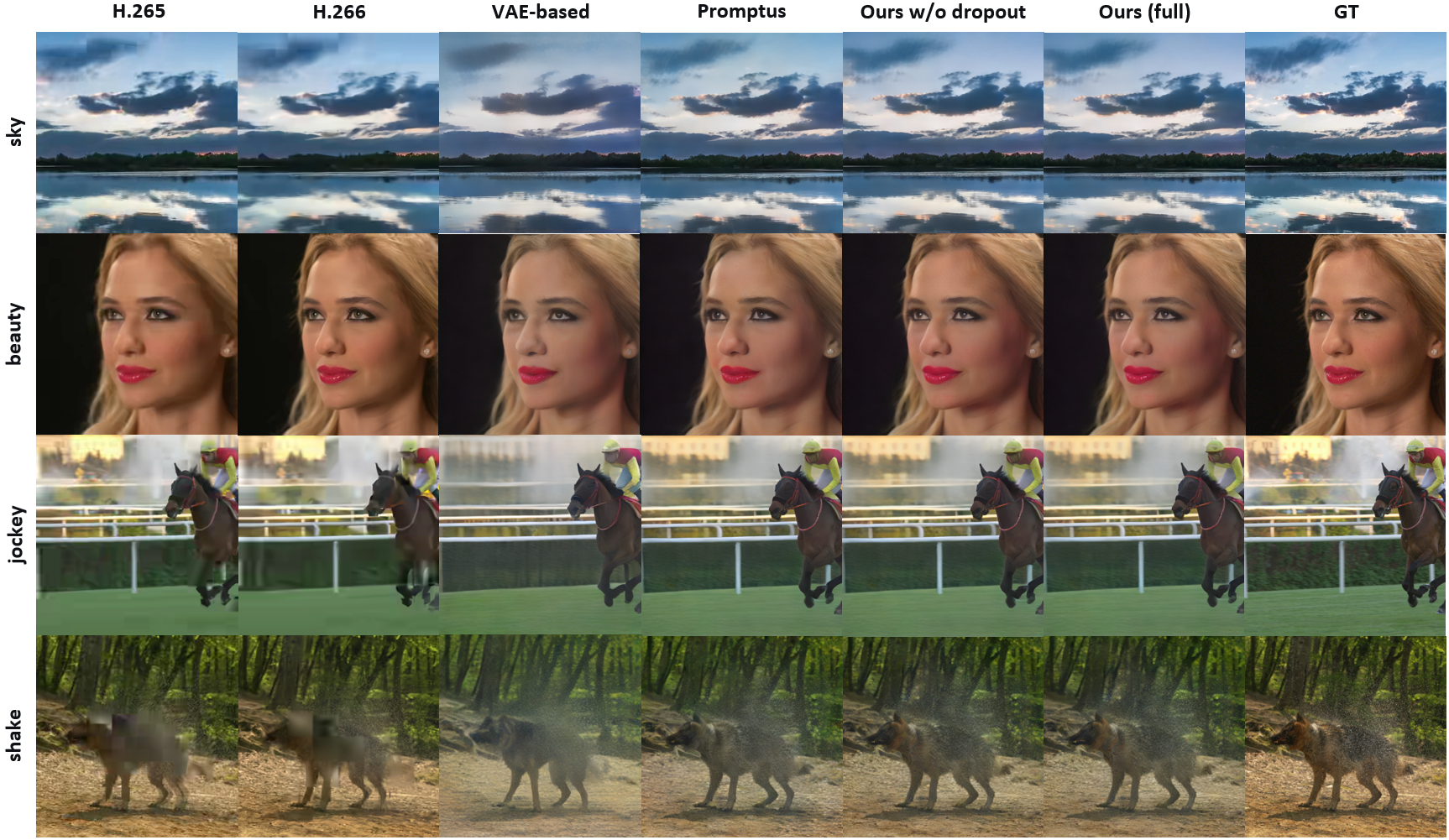}
\caption{Qualitative comparison under 70 kbps.}
\label{fig:stable_qual}
\end{figure}

Figure~\ref{fig:stable_qual} shows example frames at a low bitrate. Traditional codecs perform reasonably on videos with gentle motion, but suffer from large blocky artifacts and blur on videos with rapid motion. Ours w/o dropout produces frames with more accurate colors and finer semantic details than Promptus. Ours (full) exhibits slightly softer details due to the dropout trade-off, yet the overall visual quality remains competitive.

\subsection{4.5 Ablation Studies}

To clearly reveal the contribution of each component, we conduct ablation experiments at an extremely low prompt rank of $r = 2$ (18 kbps) on a representative video from QST, where the limited information budget magnifies quality differences across configurations.

\begin{figure}[t]
\centering
\includegraphics[width=1.0\columnwidth]{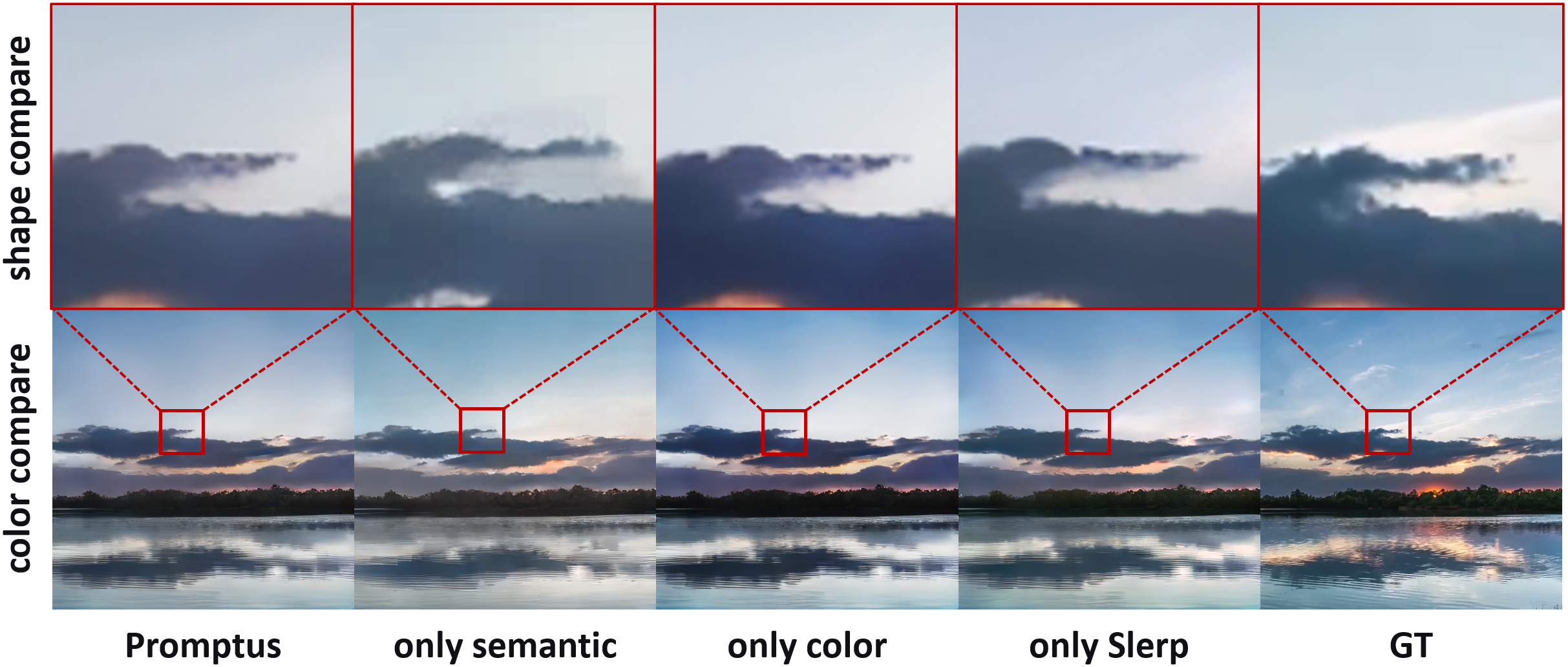}
\caption{Qualitative ablation of individual components.}
\label{fig:ablation_qual}
\end{figure}

Starting from the Promptus baseline, we evaluate the effect of individually enabling each proposed component. Table~\ref{tab:ablation} reports the results. Figure~\ref{fig:ablation_qual} visualizes the effect of each component on a representative frame. Adding the semantic similarity loss sharpens object contours and improves structural coherence, though a color shift may appear. Enabling the color loss brings the overall color distribution noticeably closer to the ground truth. With Slerp, intermediate (non-keyframe) frames exhibit fewer local artifacts and blurry regions. \textbf{Additional ablation studies (keyframe interval, loss weights) are provided in the supplementary material.}

\begin{table}[t]
\centering
\setlength{\tabcolsep}{3pt}
\caption{Ablation study. Each row enables one component on top of the Promptus baseline. (mean $\pm$ standard)}
\label{tab:ablation}
\begin{tabular}{lccc}
\toprule
Configuration & LPIPS$\downarrow$ & SSIM$\uparrow$ & PSNR$\uparrow$ \\
\midrule
Promptus       & $0.2468 \pm 0.0257$ & $0.7570 \pm 0.0425$ & $21.43 \pm 1.08$ \\
$+ \mathcal{L}_{\mathrm{sem}}$ & \textbf{$0.2294 \pm 0.0250$} & $0.7552 \pm 0.0419$ & $21.25 \pm 1.09$ \\
$+ \mathcal{L}_{\mathrm{color}}$ & $0.2346 \pm 0.0243$ & $0.7500 \pm 0.0440$ & \textbf{$21.50 \pm 1.26$} \\
$+ \mathrm{Slerp}$ & $0.2335 \pm 0.0247$ & \textbf{$0.7636 \pm 0.0430$} & $21.36 \pm 1.18$ \\
\bottomrule
\end{tabular}
\end{table}

\section{5~~~Conclusion}

We present ScalablePromptus, an enhanced prompt streaming framework that addresses two key limitations of the original Promptus. By introducing semantic and geometric prompt enhancement, we achieve modest but consistent quality gains. The core contribution---dropout training for rank-ordered representations---enables the model to gracefully tolerate truncation and packet loss \textbf{without any receiver-side adaptation}. Under lossy conditions, our method reduces the performance degradation by $82\%$--$95\%$, making prompt-based video streaming viable for real-world deployment.

\textbf{Two limitations point to directions for future work.} First, our dropout training concentrates the most critical information in the leading prompt dimensions. While this rank ordering is key to our method's robustness, it also means that these leading dimensions must be reliably delivered; in practice, a small amount of forward error correction redundancy suffices to protect them, and the overhead is acceptable given their small proportion of the total prompt size. Second, we adopt a simple Stable Diffusion model to keep training tractable and receiver-side generation fast. Larger and more capable generative models carry stronger visual priors; leveraging them could potentially achieve even higher compression ratios, and exploring this direction is a next step.


\bibliography{aaai2027}

\appendix
\section{A~~~Supplementary Materials}

\subsection{A.1 Packet Loss Resilience Under Different Channel Conditions}

Section~4.3 of the main text evaluates under one specific Gilbert-Elliott (GE) burst loss configuration ($p_{\mathrm{gb}} = 0.1$, $p_{\mathrm{bg}} = 0.3$). The GE model captures the temporal correlation of wireless packet losses through two transition probabilities: $p_{\mathrm{gb}}$ controls the probability of transitioning from a good to a bad channel state, and $p_{\mathrm{bg}}$ controls the reverse. Different $(p_{\mathrm{gb}}, p_{\mathrm{bg}})$ pairs correspond to qualitatively different loss patterns---mild channels with sporadic isolated losses, heavy channels with prolonged bursts, and severe channels where the channel is predominantly in the bad state. Here we extend the evaluation to four representative GE configurations spanning this spectrum, all at 140~kbps ($r = 16$). Table~\ref{tab:ge_supp} reports LPIPS at average loss rates of 5\%, 10\%, 15\%, and 20\%.

We restrict this comparison to Promptus and ScalablePromptus because traditional codecs (H.265, H.266, VAE) do not admit a partial-reception paradigm: a partially received coded bitstream is typically undecodable, and their packet loss resilience relies on orthogonal mechanisms such as error concealment and forward error correction. Including them in the GE table would neither be technically meaningful nor reflect their actual error-resilience capabilities. Across all channel types and loss levels, our method consistently outperforms Promptus once packet loss exceeds 0\%, confirming that the dropout-trained rank ordering confers robust loss resilience that is not tied to any particular channel model.

\begin{table}[]
\centering
\caption{LPIPS under different Gilbert-Elliott channel parameters. All at 140~kbps ($r = 16$).}
\label{tab:ge_supp}
\begin{tabular}{lccccc}
\toprule
Method & $5\%~loss$ & $10\%~loss$ & $15\%~loss$ & $20\%~loss$ \\
\midrule
\multicolumn{5}{c}{$p_{\mathrm{gb}} = 0.10$, $p_{\mathrm{bg}} = 0.50$ (mild burst)} \\
Promptus  & 0.3328 & 0.3798 & 0.3707 & 0.4418 \\
Ours & \textbf{0.3214} & \textbf{0.3255} & \textbf{0.3259} & \textbf{0.3363} \\
\midrule
\multicolumn{5}{c}{$p_{\mathrm{gb}} = 0.10$, $p_{\mathrm{bg}} = 0.30$ (moderate burst)} \\
Promptus  & 0.3345 & 0.3607 & 0.3735 & 0.4491 \\
Ours & \textbf{0.3214} & \textbf{0.3276} & \textbf{0.3289} & \textbf{0.3417} \\
\midrule
\multicolumn{5}{c}{$p_{\mathrm{gb}} = 0.20$, $p_{\mathrm{bg}} = 0.20$ (heavy burst)} \\
Promptus  & 0.3702 & 0.4143 & 0.4455 & 0.5098 \\
Ours & \textbf{0.3215} & \textbf{0.3336} & \textbf{0.3434} & \textbf{0.3536} \\
\midrule
\multicolumn{5}{c}{ $p_{\mathrm{gb}} = 0.30$, $p_{\mathrm{bg}} = 0.10$ (severe burst)} \\
Promptus  & 0.3738 & 0.4323 & 0.4603 & 0.5645 \\
Ours & \textbf{0.3236} & \textbf{0.3377} & \textbf{0.3502} & \textbf{0.3630} \\
\bottomrule
\end{tabular}
\end{table}

\subsection{A.2 Ablation on Keyframe Interval}

The keyframe interval $K$ controls the spacing at which full prompt factors are transmitted and inverted; intermediate frames are recovered via Slerp interpolation between adjacent keyframes. This creates a direct trade-off: smaller $K$ means more keyframes per second, providing denser supervision and easier interpolation, but proportionally increases the total bitrate. Conversely, larger $K$ reduces bandwidth consumption but widens the gap between keyframes, forcing Slerp to bridge longer temporal distances where the linear geodesic assumption on the hypersphere may not hold as well, leading to degraded intermediate-frame quality.

Table~\ref{tab:interval} reports LPIPS, SSIM, PSNR, and relative bitrate overhead for $K \in \{5, 10, 15, 20, 25\}$ at a fixed prompt rank of $r = 16$. The bitrate scales inversely with $K$: $K = 5$ doubles the keyframe rate (and thus the bandwidth) relative to $K = 10$, while $K = 25$ reduces it to 40\%. The quality metrics reveal diminishing returns below $K = 10$, where the bandwidth cost escalates rapidly for only marginal quality improvement. Above $K = 15$, quality degradation accelerates as interpolation errors accumulate. We select $K = 10$ (3 keyframes per second) as the default in our main experiments, providing a practical operating point that balances reconstruction fidelity with transmission efficiency.

\begin{table}[]
\centering
\caption{Ablation on keyframe interval ($r = 16$).}
\label{tab:interval}
\begin{tabular}{lcccc}
\toprule
Interval & Bitrate & LPIPS$\downarrow$ & SSIM$\uparrow$ & PSNR$\uparrow$ \\
\midrule
5   & 281.86 kbps & 0.2692 & 0.7296 & 28.69 \\
10  & 140.93 kbps & 0.2793 & 0.7226 & 28.34 \\
15  & 93.95 kbps & 0.2836 & 0.7214 & 28.26 \\
20  & 70.46 kbps & 0.2907 & 0.7144 & 28.04 \\
25  & 56.37 kbps & 0.3020 & 0.7099 & 27.88 \\
\bottomrule
\end{tabular}
\end{table}

\subsection{A.3 Ablation on Loss Weights}

In the main experiments ($r = 8$ or higher), the prompt representation has sufficient capacity to simultaneously satisfy pixel-level fidelity, semantic alignment, and color consistency constraints, making it difficult to isolate the contribution of each loss term. To reveal their distinct effects, we deliberately conduct this ablation at an extreme low rank of $r = 2$ (18~kbps), where the prompt carries only 2~202 scalar parameters per keyframe. In this severely constrained regime, every bit of embedding capacity is precious: the different loss terms actively compete for the limited representational budget, forcing the optimization to make explicit trade-offs. This is precisely the setting where each loss term's functional role becomes most visible---much like how a low-bandwidth scenario reveals which visual information a codec considers most essential.

\begin{figure}[t]
\centering
\includegraphics[width=1.0\columnwidth]{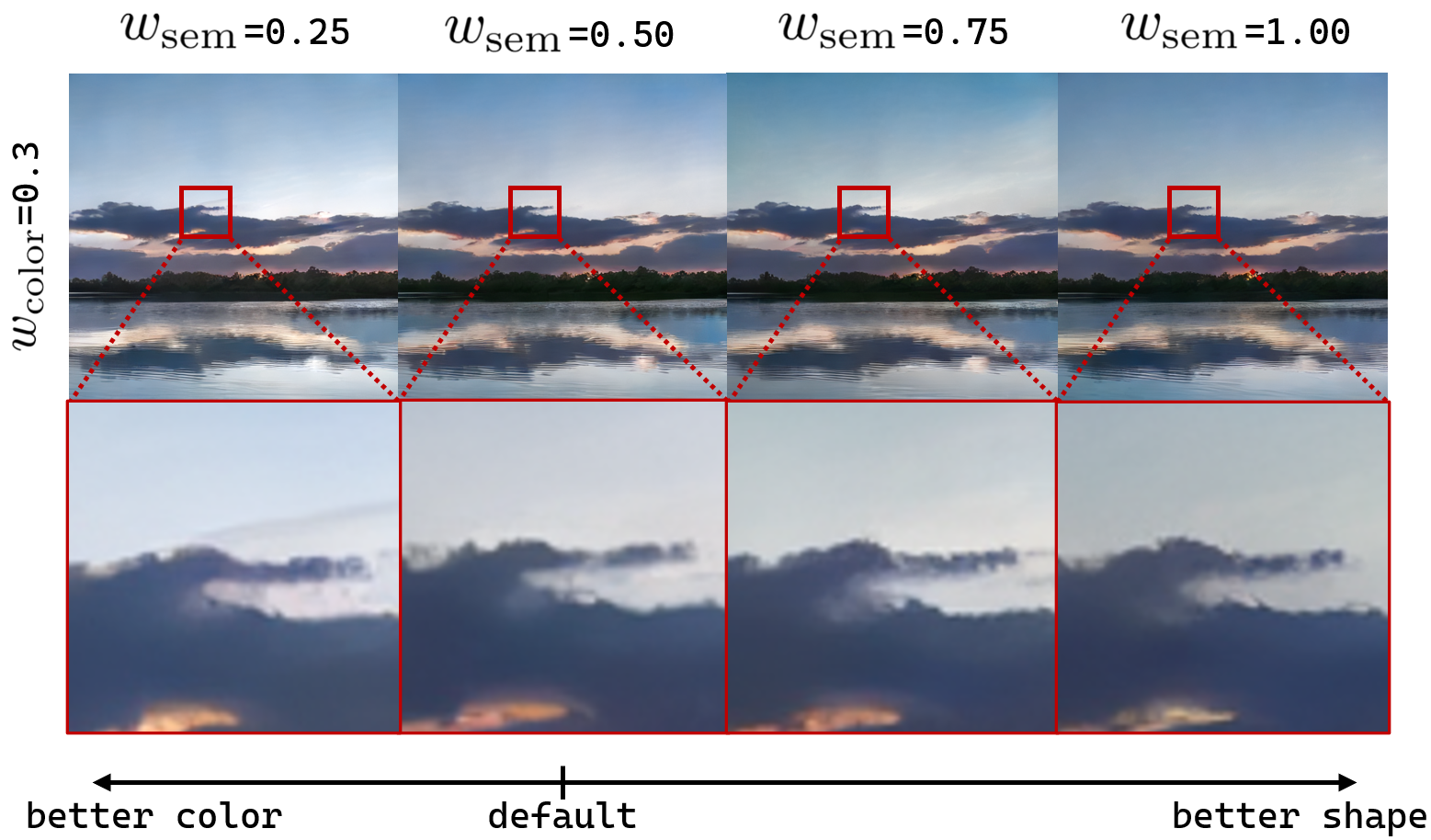}
\caption{Qualitative ablation of $w_{\mathrm{sem}}$ ($w_{\mathrm{color}} = 0.3$ fixed) at $r = 2$. Left to right: $w_{\mathrm{sem}} = 0.25, 0.50, 0.75, 1.00$. Increasing semantic weight sharpens structure but gradually desaturates color.}
\label{fig:loss_sem}
\end{figure}

\begin{figure}[t]
\centering
\includegraphics[width=1.0\columnwidth]{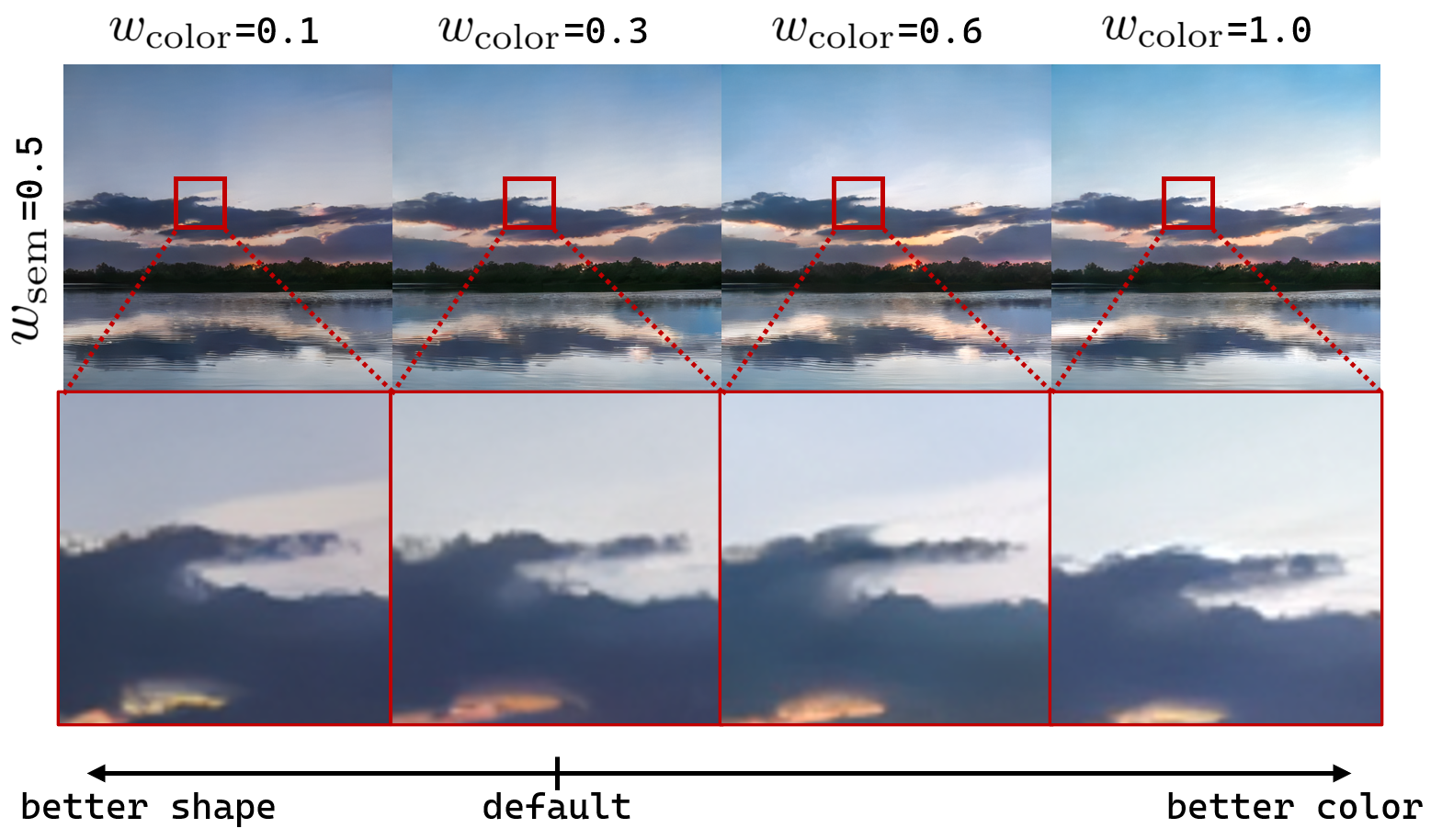}
\caption{Qualitative ablation of $w_{\mathrm{color}}$ ($w_{\mathrm{sem}} = 0.50$ fixed) at $r = 2$. Left to right: $w_{\mathrm{color}} = 0.1, 0.3, 0.6, 1.0$. Increasing color weight improves color fidelity but blurs fine details.}
\label{fig:loss_color}
\end{figure}

Table~\ref{tab:loss_weights} reports LPIPS under two sweeps. In the first sweep (top), we vary $w_{\mathrm{sem}}$ while keeping $w_{\mathrm{color}} = 0.3$ fixed; in the second sweep (bottom), we vary $w_{\mathrm{color}}$ while keeping $w_{\mathrm{sem}} = 0.5$ fixed. Figure~\ref{fig:loss_sem} and \ref{fig:loss_color} provide qualitative visualizations of representative frames at selected weight combinations.

Two clear and opposing trends emerge. \textbf{Increasing $w_{\mathrm{sem}}$} progressively strengthens structural and semantic alignment: object contours become sharper, and the overall layout adheres more faithfully to the ground truth. However, as semantic supervision dominates, the optimization allocates less of the limited prompt capacity to color information, and the generated frames gradually desaturate, trending toward a grayish appearance. \textbf{Increasing $w_{\mathrm{color}}$} progressively improves color fidelity: the per-channel mean and variance of the generated frame converge toward those of the ground truth, producing more vivid, natural tones. Yet with color alignment consuming a larger share of the prompt budget, fine-grained structural details---particularly around object boundaries and high-frequency texture regions---become increasingly blurred.

The default weights $w_{\mathrm{sem}} = 0.5$, $w_{\mathrm{color}} = 0.3$ used throughout the main text reside at the intersection of these two curves, representing a calibrated compromise where neither structural integrity nor color naturalness is unduly sacrificed.

\begin{table}[t]
\centering
\setlength{\tabcolsep}{8pt}
\caption{LPIPS under varying loss weights ($r = 2$).}
\label{tab:loss_weights}
\begin{tabular}{lcccc}
\toprule
\multicolumn{5}{c}{\textit{Varying $w_{\mathrm{sem}}$ ($w_{\mathrm{color}} = 0.3$ fixed)}} \\
$w_{\mathrm{sem}}$ & 0.25 & 0.50 (default) & 0.75 & 1.00\\
\midrule
LPIPS$\downarrow$ & 0.2752 & \textbf{0.2640} & 0.2704 & 0.2695  \\
\midrule
\multicolumn{5}{c}{\textit{Varying $w_{\mathrm{color}}$ ($w_{\mathrm{sem}} = 0.50$ fixed)}} \\
$w_{\mathrm{color}}$ & 0.1 & 0.3 (default) & 0.6 & 1.0 \\
\midrule
LPIPS$\downarrow$ & 0.2763 & \textbf{0.2640} & 0.2733 & 0.2745 \\
\bottomrule
\end{tabular}
\end{table}

\subsection{A.4 On the Choice of Bitrate Range}

Our stable-network experiments (Sec~4.4) focus on the 70--140~kbps range. We think that this is precisely the regime where prompt-based generative streaming offers the most compelling and practical advantages. At high bitrates (e.g., above 300~kbps), modern codecs such as H.266 already achieve excellent perceptual quality---blocking and blurring artifacts become nearly imperceptible, and the room for further compression via generative priors narrows considerably. The additional complexity of training and deploying a diffusion-based decoder is harder to justify when a standard codec already delivers satisfactory results.

In contrast, the low-bitrate regime (roughly 50--200~kbps) exposes a fundamental limitation of traditional codecs: constrained by the Shannon rate-distortion bound, they are forced to aggressively quantize transform coefficients, discarding perceptually critical high-frequency information. This manifests as severe blocking, blurring, and texture loss that no amount of encoder tuning can avoid---the information has been irreversibly discarded. Generative methods resolve this tension through a fundamentally different mechanism: rather than compressing and discarding pixel-level detail, they transmit a compact semantic representation and rely on a powerful generative prior to reconstruct plausible, high-frequency content at the receiver. The transmitted bitstream encodes \emph{what the scene contains} rather than \emph{what each pixel looks like}, decoupling bitrate from pixel count.

The 70--140~kbps range is representative of bandwidth-constrained scenarios such as congested cellular networks, multi-user WiFi streaming, and low-power IoT video links, where traditional codecs exhibit their most objectionable artifacts. This is also where our method's robustness gains---82\%--95\% reduction in truncation-induced degradation---deliver the greatest practical impact, making prompt-based streaming not just theoretically interesting but deployment-ready for challenging real-world conditions.

\end{document}